\documentclass[10pt,twocolumn,letterpaper]{article}

\usepackage{cvpr}
\usepackage{times}
\usepackage{epsfig}
\usepackage{graphicx}
\usepackage{amsmath}
\usepackage{amssymb}
\usepackage[none]{hyphenat}
\usepackage{multirow}
\usepackage{rotating}

\usepackage[pagebackref=true,breaklinks=true,letterpaper=true,colorlinks,bookmarks=false]{hyperref}

\begin{document}

\title{$i$CTGAN--An Attack Mitigation Technique for Random-vector Attack on Accelerometer-based Gait Authentication Systems}

\author{Jun Hyung Mo\\
Haverford College, USA\\
{\tt\small junmo318@gmail.com}
\and
Rajesh Kumar\\
Bucknell University, USA\\
{\tt\small rajesh.kumar@bucknell.edu}
}

\maketitle

\begin{abstract}
A recent study showed that commonly (vanilla) studied implementations of accelerometer-based gait authentication systems ($v$ABGait) are susceptible to random-vector attack. The same study proposed a beta noise-assisted implementation ($\beta$ABGait) to mitigate the attack. In this paper, we assess the effectiveness of the random-vector attack on both $v$ABGait and $\beta$ABGait using three accelerometer-based gait datasets. In addition, we propose $i$ABGait, an alternative implementation of ABGait, which uses a Conditional Tabular Generative Adversarial Network. Then we evaluate $i$ABGait's resilience against the traditional zero-effort and random-vector attacks. The results show that $i$ABGait mitigates the impact of the random-vector attack to a reasonable extent and outperforms $\beta$ABGait in most experimental settings. \footnote{© 2022 IEEE. Personal use of this material is permitted. Permission from IEEE must be obtained for all other uses, in any current or future media, including reprinting/republishing this material for advertising or promotional purposes, creating new collective works, for resale or redistribution to servers or lists, or reuse of any copyrighted 
component of this work in other works. To appear in IEEE International Joint Conference on Biometrics (IJCB 2022), Oct.10-13, 2022, Abu Dhabi, United Arab Emirates.}


\end{abstract}
\vspace{-0.25in}
\section{Introduction} 
\subsection{Gait}
Gait refers to how a person walks. If captured properly, it can be used to identify or verify one's identity \cite{Ross2018, WearableSensorGaitSurvey}. While Aristotle studied it in $350$ BC, human gait became popular in the biometrics community in the 1990s primarily due to the advancements in the means (\eg, cameras, force plates, 3D motion detection) that could capture gait more precisely than ever before. Later, with the advent and massive adaptation of smart devices, especially smartphones and smartwatches, the study of gait attracted even more researchers. A smart device consists of inertial sensors, \eg, accelerometer, gyroscope, magnetometer, and rotation vector, that help capture and characterize gait in ways not studied before. For example, an accelerometer captures and characterizes gait in terms of acceleration, and a gyroscope does the same in rotation. The gait captured via accelerometer and gyroscope is considered the most distinguishing among individuals \cite{WearableSensorGaitSurvey, PhoneMovementPrinceton}. Most gait recording technologies can be grouped into intrusive or constrained (cameras and force plates are used) and non-intrusive (personal devices such as smartphones and smartwatches). Different recording mechanisms offer different application possibilities, such as using gait for surveillance, identification, authentication, and healthcare monitoring.  

In this paper, we focus on Accelerometer-based Gait Authentication Systems (ABGait) \cite{WearableSensorGaitSurvey,PhoneMovementPrinceton}. ABGait has attracted the attention of defense organizations like the Defense Advanced Research Projects Agency and The Defense Information Systems Agency of the United States. These agencies are in the process of implementing security measures that will utilize a variety of behavioral biometrics, including ABGait \cite{DISA}. ABGait presumably offers immunity to shoulder surfing, social engineering, or spoof-based circumvention attempts \cite{Muaaz2017} in addition to convenience due to its passive and non-intrusive nature.

\subsection{Accelerometer-based Gait Authentication}
Previous studies have demonstrated that inertial sensor-based gait patterns, regardless of whether the sensors were embedded into a smartphone or smartwatch, are sufficiently unique for individuals under different device usage contexts \cite{Ross2018, WearableSensorGaitSurvey}. Once the correct device-usage context is identified using state-of-the-art human activity recognition systems, gait recordings are segmented into individual gait cycles or fixed-length frames (often overlapping). Gait cycles are extracted by identifying minima, maxima, zero crossing, or conducting phase analysis. On the other hand, fixed-length frames are created using a sliding window-based mechanism. The frames of $8$--$12$ seconds with sliding length half have achieved superior results \cite{PhoneMovementPrinceton, KumarArm}. It is worth noting that the frames of such lengths consist of multiple gait cycles. For cycle-based segmentation, a distance-based metric is usually applied to establish a match threshold for classifying each cycle as genuine, or an impostor \cite{Muaaz2017,2007GafurovAttackGender}. Recently researchers have focused on the extraction of statistical and frequency domain features from the cycles and the application of machine learning algorithms. On the other hand, researchers have established a machine learning pipeline to train the authentication models for the fixed-length segmentation case. \cite{PhoneMovementPrinceton, KumarArm}. 

Authentication models trained using the fixed-length frame with a machine learning pipeline have usually outperformed cycle-based approaches with distance metrics, \eg, Euclidean or Manhattan distances or distance measures such as Dynamic Time Warping (DTW) \cite{FrameIsBetterThanCycleInGait, kumar2020, DistanceBasedVsMachineLearning}. An experiment on the same dataset showed a difference of $13.6$\% \cite{DistanceBasedVsMachineLearning} with the frame-based training achieving $92.7$\% accuracy and cycle-based approach achieving $79.1$\%. Recent studies with a better machine learning pipeline have achieved error rates under $5$\% \cite{PhoneMovementPrinceton}. 

Because authentication systems play an essential role in securing users' data and privacy, it is paramount that the authentication systems can defend themselves from adversarial attacks. Zhao \etal \cite{Zhao2020} examined the vulnerability of ABGait from random-vector attacks. The performance of ABGait is generally evaluated in terms of the number of false accepts (or false positives) and false rejects (false negatives). The lower these numbers are, the better the ABGait is. However, Zhao \etal \cite{Zhao2020} pointed out that the false-positive rates of ABGait could be misleading as they are computed entirely based on the impostor samples that are available at the time of testing.

Interestingly, until Zhao \etal \cite{Zhao2020}, researchers had overlooked the possibility of someone attacking ABGait with random feature vectors with the assumption that the attacker would know the length of the feature vectors that are used to train the ABGait. Zhao \etal \cite{Zhao2020} showed that ABGait accepts even uniform random inputs. Additionally, they found that the probability of random vector acceptance is much higher than the false-positive rates. They suggested that the acceptance and rejection regions created during the ABGait training are much bigger than the false accept and false reject rates. Therefore the performance evaluation of ABGait should include both measures, i.e., False Acceptance Rates (FAR) and Acceptance Region (AR), to assess the ability of ABGait to defend itself from active adversarial attempts such as random-vector attacks \cite{Zhao2020}. 

\subsection{Adversarial Scenarios} Its commonly believed that behavioral biometrics such as ABGait require more effort to circumvent than physical biometrics. However, some studies have pointed out that ABGait is vulnerable to treadmill-assisted imitation attacks \cite{kumar2020} and random-vector attacks \cite{Zhao2020}. These two attack paradigms are substantially different. The former requires minimal system knowledge, is difficult to launch, and requires artifacts such as a treadmill. In contrast, the latter merely requires the generation of random vectors. However, it also assumes that the attacker would have access to the authentication API and the knowledge of feature space (length of feature vectors, normalization methods, etc.) \cite{Zhao2020}. Considering the feasibility and ease with which a random vector attack can be executed, we focus on the same in this study. 

\subsection{Possible Countermeasures}
The possible countermeasures to the random-vector attack include training the authentication models using synthetic data. Zhao \etal \cite{Zhao2020} used synthetic noise generated around genuine samples to train the authentication models. The generated noise around the genuine samples was labeled as impostors during the training process to increase the impostor samples' variance. The model showed greater resilience to random-vector attacks than the models without synthetic data-based training. Quantitatively, the models achieved a reduced area of the acceptance region, representing a potentially errant input as genuine. The idea worked on the dataset \cite{HAR} that was studied in \cite{Zhao2020} but needs to be tested on multiple datasets, including ones collected from smartwatches. Inspired by the previous countermeasure, we propose using Conditional Tabular Generative Adversarial Networks (CTGAN) \cite{CTGANOriginal} on impostor samples to increase the variance of the impostor samples and test its usefulness in mitigating the random-vector attack while maintaining the reported performance. GAN has been successfully used to safeguard gait-based key generation from vision-based side-channel attacks in the past \cite{GANBasedDefenseVisionBased}. 

\subsection{Main Contributions}
First, we implemented vanilla ABGait ($v$ABGait) with no mitigation technique in the pipeline. Then we tested the same on three different datasets consisting of a different number of users, samples per user, and feature set and computed the performance measures, i.e., False Accept Rate (FAR), False Reject Rate (FRR), Acceptance Region (AR), and Half Total Error Rate (HTER) under the zero-effort and random-vector attack scenarios. Second, we included the mitigation technique proposed in Zhao \etal \cite{Zhao2020} which resulted in ($\beta$ABGait), and evaluated the same on all three datasets using the aforementioned metrics under the zero-effort and random-vector attack scenarios. Third, we introduced $i$CTGAN, a CTGAN-assisted impostor samples generator, in the ABGait training pipeline, evaluated its performance on the three datasets, and compared its performance with $\beta$ABGait and $v$ABGait. The pre-processed dataset and code is available at \cite{CodeDataset}. 


\section{Related Work}
\label{RelatedWorks}
Zhao \etal \cite{Zhao2020} is the most closely related work to this paper. The authors theoretically and experimentally demonstrated that for machine learning-based ABGait, the acceptance region is significantly larger than the false acceptance rate (FAR). Consequently, an attacker with access to ABGait via a black-box feature vector Application Programming Interface (API) can gain authorized access to the protected system by supplying randomly generated vectors. On average (over the four classifiers studied), the likelihood of an attacker being successful by supplying feature vectors with uniform random values (random-vector attack) was found to be much higher (AR:$16.25$\%) than the false accept rate (FAR:$14.5$\%) for the dataset studied in the paper. The gap between AR and FAR was even bigger in the case of touch stroke-based authentication systems \cite{Zhao2020}. 

A major limitation of the study by Zhao \etal \cite{Zhao2020} is that the idea is tested only on one gait dataset. We apply the attack and the countermeasure to three different datasets to assess the random-vector attack's impact and proposed countermeasure presented by \cite{Zhao2020}. In addition, we proposed an alternative countermeasure based on CTGAN, which achieved superior results in most experimental setups (datasets and classification algorithms). We note that the countermeasures proposed in this paper and Zhao \etal \cite{Zhao2020} are tested against random-vector attacks only. In the future, we aim to test these ideas against cluster-based attacks \cite{Zhu2021} and treadmill-assisted attacks \cite{kumar2020} among others. A more comprehensive classification of possible attacks on behavioral biometrics e.g., touch gestures can be found in \cite{GANTOuchTBIOM2022}. 

\begin{figure*}[htp]
    \centering
    \includegraphics[width=6.7in, height = 2.2in]{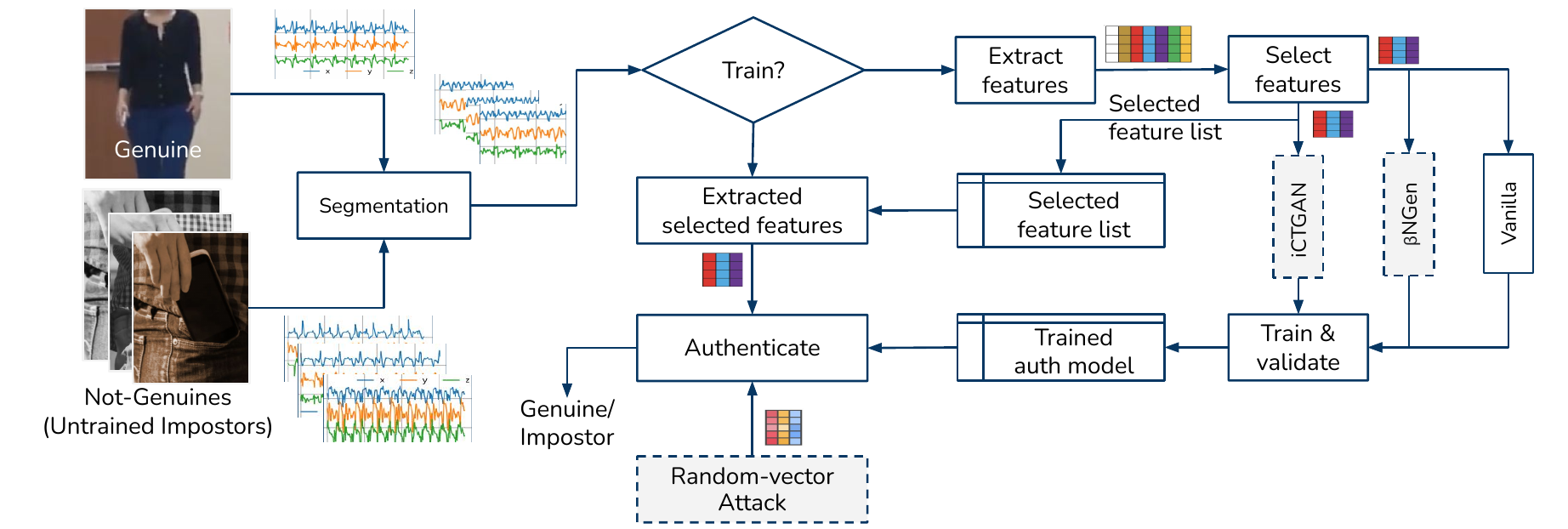}
    \caption{The system architecture of ABGait with $v$ABGait, $\beta$ABGait, and $i$ABGait variations. The novel components in the pipeline are highlighted with dashed lines and light gray background. The first is at the bottom, named random-vector attack. The second and third are the $\beta$ noise generator ($\beta$NGen) and the proposed $i$CTGAN. $\beta$NGen generates noise around genuine samples. The $\beta$NGen generated samples are considered impostors during the training. In contrast, $i$CTGAN generates more impostor samples from real impostor samples. We refer to $\beta$NGen and $i$CTGAN-based systems as $\beta$ABGait and $i$ABGait, respectively. In contrast, the ones trained without $\beta$NGen or $i$CTGAN are vanilla ABGait ($v$ABGait).}
    \label{AdversarialISGait}
\end{figure*}

\section{Design of Experiments}
\label{ExperimentalDetails}
\subsection{Datasets}
We use three datasets in our experiments. The Human Activity Recognition (HAR) dataset, collected via smartphones, was used to maintain continuity with the previous \cite{Zhao2020}. In addition, we include a smartphone-based dataset of $55$ users from a recent study \cite{kumar2020}. Moreover, we included a smartwatch-based gait dataset used in \cite{KumarArm}. The purpose of including multiple datasets and different sets of features for each dataset was to investigate whether the defense ideas proposed in this paper as well as in \cite{Zhao2020} are applicable regardless of the feature set, device, and dataset. 

\subsubsection{HAR dataset}
Anguita created the HAR (Human Activity Recognition) dataset \etal \cite{HAR}. The dataset has been widely studied to test human activity (sitting, laying down, walking, running, walking upstairs or downstairs) recognition frameworks via accelerometer and gyroscope sensors of a smartphone \cite{HAR}. The dataset contains a total of $30$ users, with an average of $343$ samples for each user. The dataset was re-purposed for authentication because the dataset contains unique ids for each user. Under the context of authentication, the different types of activity served as an additional feature. A group of $30$ volunteers aged from $19$ to $48$ years was given a waist-mounted Samsung Galaxy S II Smartphone. They performed each of the different types of activities for $15$ seconds. 

\subsubsection{Watch dataset} Kumar \etal \cite{KumarArm} collected the smartwatch dataset. It consists of $40$ participants who tied Samsung Galaxy Gear S on their wrists and walked in a corridor in two different sessions, each about two minutes. The dataset has both accelerometer and gyroscope readings. We included only accelerometer readings in this study for consistency with the HAR dataset and \cite{Zhao2020}. Thirty-four subjects were in the age range of $20$ to $30$s, four between $30$s to $40$s, and two older than $50$ years old. We performed the same preprocessing and feature extractions steps proposed in \cite{KumarArm}.  

\subsubsection{Phone dataset}
This dataset has developed over time and has been used in \cite{kumar2020, Kumar2015, MyPhDThesis}. The dataset consists of $55$ users who walked naturally for about $2-4$ minutes, keeping an HTC-One M8 smartphone in the right pocket of their pants with the screen facing the participant's body. Although the dataset consisted of readings from four sensors, we used only accelerometer sensor readings to be consistent with the previous studies and the other datasets.

\subsubsection{Synthetic dataset} In addition to the real data collected from participants, the training and testing process included synthetic data generated by using a uniform random generator, applying $\beta$ noise generator ($\beta$NGen) on genuine samples and CTGAN (CTGAN) on zero-effort impostor samples. The data generated by the uniform random generator was used to launch the random-vector attack. On the other hand, the data generated by $\beta$NGen and $i$CTGAN were used as impostor samples during the training of ABGait. 

\subsection{Feature Analysis} The steps of data preprocessing and feature engineering and analysis were replicated from previous studies which originally proposed the datasets. Keeping the steps similar to previous studies was deliberate, primarily investigating the effectiveness of the $i$CTGAN-based countermeasure proposed in this study on different types of datasets, feature sets, and classification models. In particular, the feature extraction from the HAR dataset was the same as described in \cite{Zhao2020}, while the feature extraction process from the Watch and Phone datasets was the same as detailed in \cite{KumarArm} and \cite{kumar2020} respectively.

\subsection{Choice of Classifiers}
Since this was an extension of the previous study by Zhao \etal \cite{Zhao2020}, we confined ourselves to the same classifiers used in that study. In particular, four classifiers viz. Support Vector Machine with Linear Kernel (LINSVM) and Radial Basis Kernel (RBFSVM), Random Forests (RNDF) implemented using Scikit-learn \cite{scikitpackage}, and Deep Neural Networks implemented using TensorFlow (TFDNN) \cite{tensorflow}. SVM is considered one of the finest binary classifiers for authentication purposes \cite{kumar2020,DistanceBasedVsMachineLearning,Zhu2021,ISBA2018}. RNDF on the other hand has been widely used in this domain \cite{FrameIsBetterThanCycleInGait,kumar2020,ISBA2018}. Deep Neural Network with Convolution Neural Networks (CNN) has achieved good results in this domain \cite{WearableSensorGaitSurvey, DeepCNNGait}. The use of multiple classifiers from multiple paradigms was to test the effectiveness of the random-vector attack and the countermeasures on different datasets. The hyperparameters and other implementation details were kept as suggested in \cite{Zhao2020}. 

\subsection{Mitigation Techniques}
Zhao \etal \cite{Zhao2020} explain the reason behind the success of the random attack in detail. In a nutshell, the acceptance region is much bigger than the one captured by FAR. The idea behind mitigation, therefore, is to reduce the acceptance region. The mitigation techniques studied in this paper revolve around increasing the variance of impostor samples while avoiding any overlap with the genuine samples. The $v$ABGait implementation uses feature vectors created from all the participants but the genuine user as the impostor. The number of non-genuine users for any genuine users is limited in any dataset (assuming that no dataset contains all possible users. This limitation leads to a sparse representation of the impostors, leaving a wider acceptance region. This problem can be addressed by (a) collecting large numbers of samples from a large number of users, which is often infeasible, or (b) by finding algorithmic ways to increase the variance in impostor samples. Option (b) is what Zhao \etal \cite{Zhao2020}, and we follow in this paper. Both of the methods are described in the following subsections.  

\subsubsection{$\beta$ Noise Generator ($\beta$NGen)} Originally proposed by \cite{Zhao2020}, ABGait is trained using three types of samples. The first type is genuine samples, the second type is impostor samples taken from users other than the genuine users, and the third type is the generated samples from a beta distribution dependent on the positive user. The first type was labeled genuine, while the rest were labeled impostors during the training. The sample generation process is described below. The $\beta$ distribution is parameterized by two positive shape parameters, $\alpha$, and $\beta$, which are defined as follows: \\
    $\alpha_{i} = \left | 0.5 - \mu _{i} \right | + 0.5 \text{ and } \beta_{i} = 0.5$
An impostor vector $x$ is constructed by sampling its $i$th element from one of two distributions where $\mathit{B}\varepsilon\left ( \alpha _{i}, \beta_{i}  \right )$ represents the beta distribution: \\
$\mathit{B}\varepsilon\left ( \alpha _{i}, \beta_{i}  \right ) \textrm{, if } \mu \leq  0.5 \text{ \& } 1 - \mathit{B}\varepsilon \left ( \alpha _{i}, \beta_{i}  \right ) \textrm{, if } \mu >  0.5 $
The two cases make sure that symmetric noise is added as the mean moves over to either side of $0.5.$

\subsubsection{Impostor CTGAN ($i$CTGAN)} An alternative way of increasing the variance of the impostor samples is to generate more impostor samples via the widely popular GAN \cite{GAN}. However, ABGait uses handcrafted features of mixed data types (\eg, \# of peaks vs. avg acceleration) with the non-necessarily-normal distribution. GAN was to likely suffer in this case \cite{CTGANOriginal} which prompted us to utilize Conditional Tabular Generative Adversarial Networks (CTGAN) \cite{CTGANOriginal} which addresses the issues of GAN with tabular data. In particular, CTGAN uses mode-specific normalization to address the weakness of GAN with the non-normality and multi-modal distribution. CTGAN also leverages recent advances in GAN training, such as the loss function proposed in \cite{WGANGP} and the Discriminator architecture suggested in PacGAN \cite{PacGAN}, which improves both the training stability and quality of the generated data. 

Let impostor feature matrix be represented as a table $T$ that contains $N_{c}$ continuous, and $N_{d}$ discrete feature columns (random variables). These feature variables form an unknown joint distribution $\mathbb{P}(C_{1:N_{c}},D_{1:N_{d}}$). Let $r_j = \{c_{1,j}, ...,  c_{N_c,j} , d_{1,j} , . . . , d_{N_d,j}\}, j \in \{1, . . . , n\}$ represents one sample. The objective of the synthetic data generator $\mathbb{G}$ is to learn from the table $T$ and generate $T'$. $T'$ is assessed by a discriminator or critic $\mathbb{C}$ which estimates the distance between the learned conditional distribution $\mathbb{P}_G(r_j|cond)$ and the conditional distribution on real data $\mathbb{P}(r_j|cond)$. The sampling of real training data and the construction of $cond$ vector should comply to help $\mathbb{C}$ estimate the distance. The construction process of $cond$ is explained in \cite{CTGANOriginal}. 

\begin{figure*}[htp]
  \centering
        \begin{subfigure}[b]{0.99\textwidth}
                 \includegraphics[width=6.8in, height=1.21in]{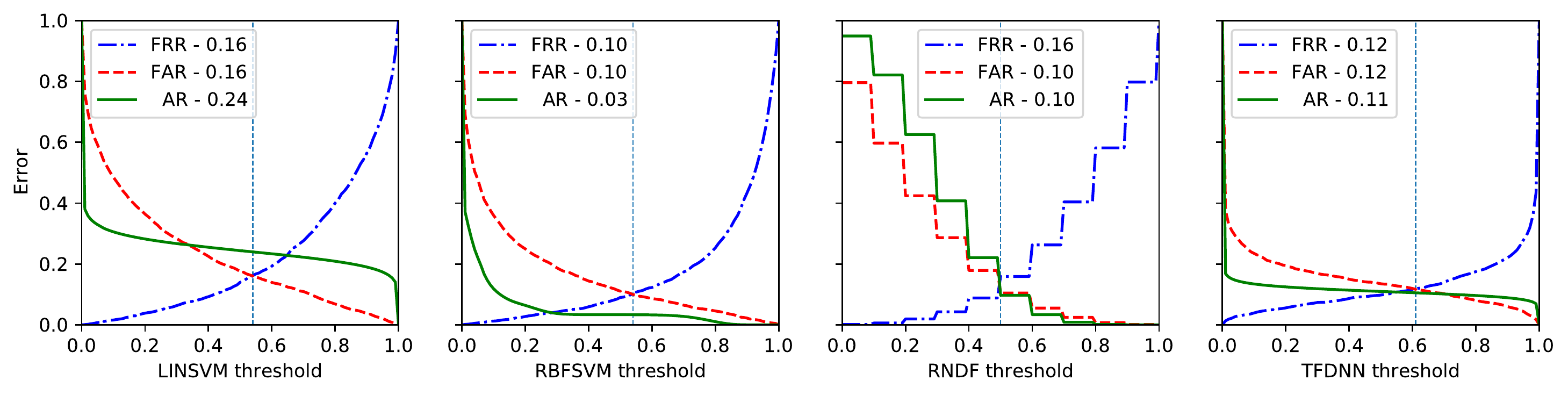}
                 \caption{The performance of \textbf{$v$ABGait} under both Zero-effort (FAR) and Random-vector attack (AR).}
                 \label{v-HAR}
         \end{subfigure}
         
         \begin{subfigure}[b]{0.99\textwidth}
                 \includegraphics[width=6.8in, height=1.21in]{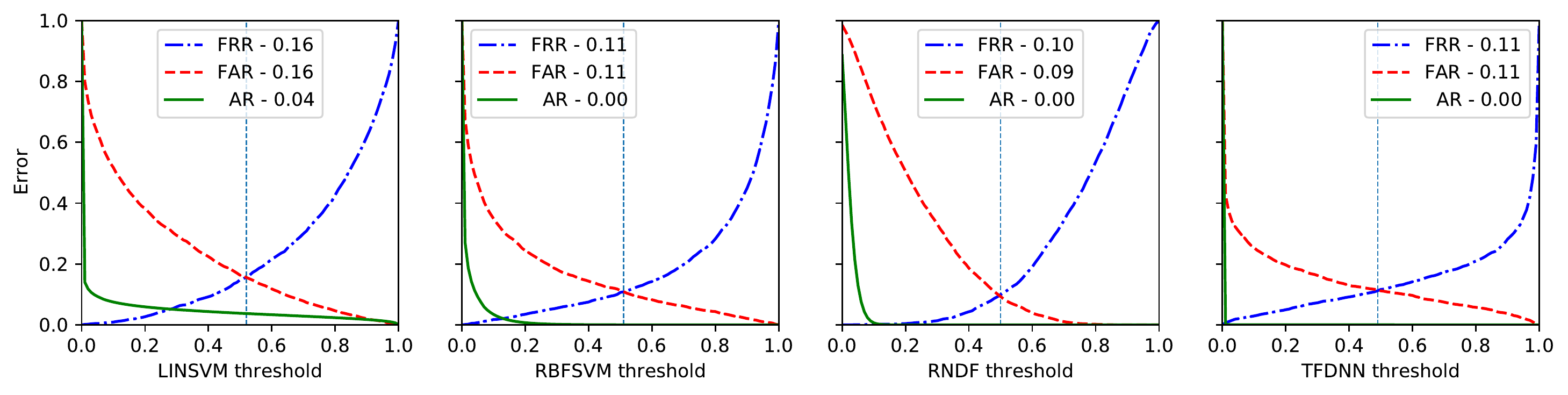}
                 \caption{The performance of \textbf{$\beta$ABGait} under both Zero-effort (FAR) and Random-vector attack (AR).}
                 \label{b-HAR}
         \end{subfigure}
         
         \begin{subfigure}[b]{0.99\textwidth}
                 \includegraphics[width=6.8in, height=1.21in]{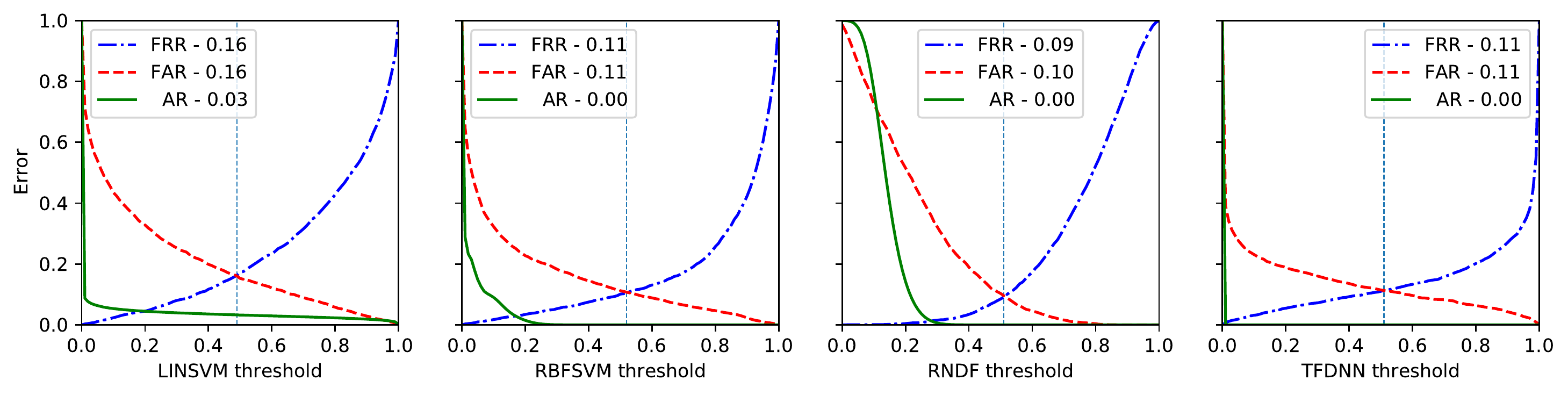}
                 \caption{The performance of \textbf{$i$ABGait} under both Zero-effort (FAR) and Random-vector attack (AR).}
                 \label{i-HAR}
         \end{subfigure}
    \caption{The performance of different \textit{ABGait} under different attack scenarios on \textbf{HAR dataset} \cite{HAR}.}
    \label{HARResults}
\end{figure*}


\begin{figure*}[htp]
  \centering
         \begin{subfigure}[b]{0.99\textwidth}
                 \includegraphics[width=6.8in, height=1.21in]{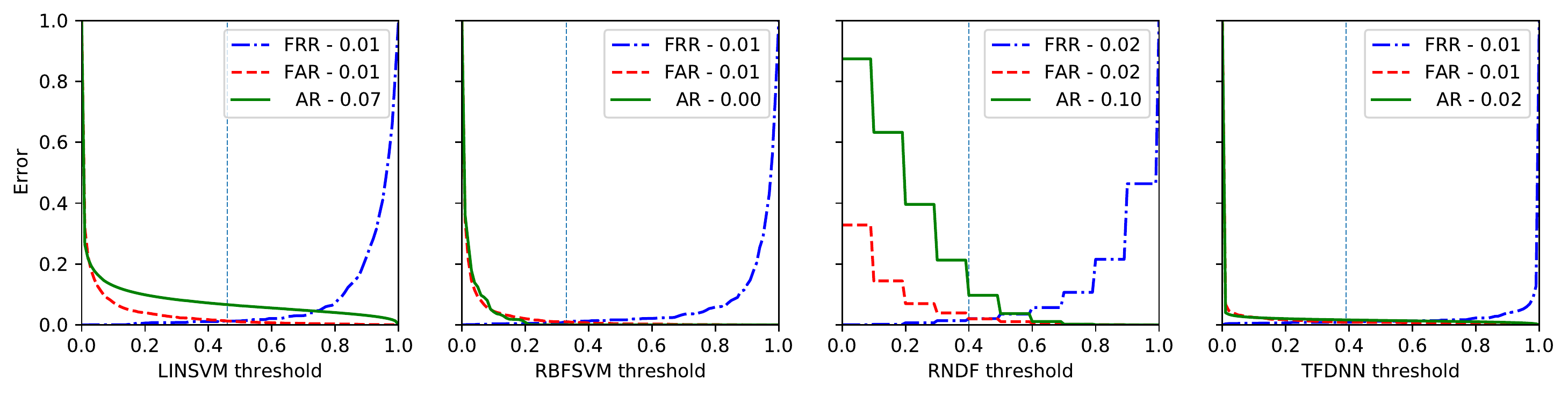}
                 \caption{The performance of \textbf{$v$ABGait} under both Zero-effort (FAR) and Random-vector attack (AR).}
                 \label{v-Phone}
         \end{subfigure}
         
         \begin{subfigure}[b]{0.99\textwidth}
                 \includegraphics[width=6.8in, height=1.21in]{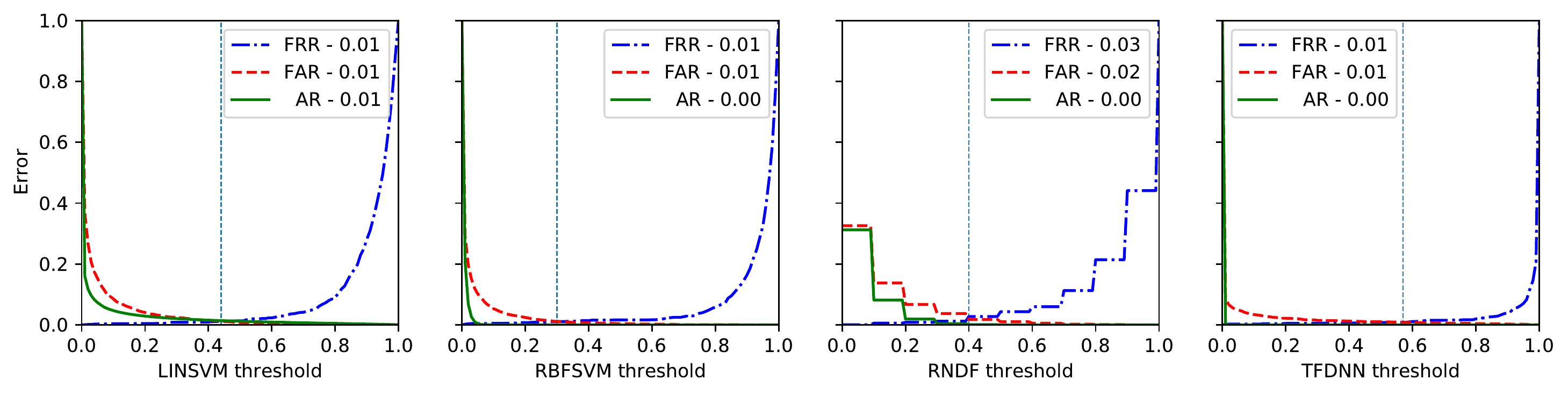}
                 \caption{The performance of \textbf{$\beta$ABGait} under both Zero-effort (FAR) and Random-vector attack (AR).}
                 \label{b-Phone}
         \end{subfigure}
         
         \begin{subfigure}[b]{0.99\textwidth}
                 \includegraphics[width=6.8in, height=1.21in]{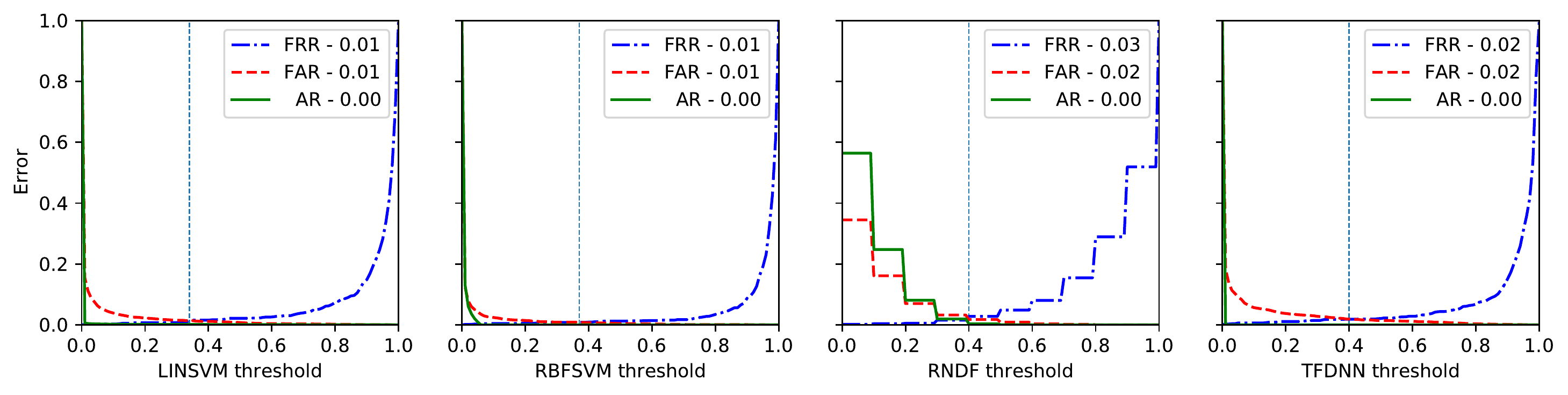}
                 \caption{The performance of \textbf{$i$ABGait} under both Zero-effort (FAR) and Random-vector attack (AR).}
                 \label{i-Phone}
         \end{subfigure}
    \caption{The performance of different \textit{ABGait} under different attack scenarios on the \textbf{Phone dataset}  \cite{kumar2020}.}
    \label{PhoneResults}
\end{figure*}


\begin{figure*}[htp]
  \centering
        \begin{subfigure}[b]{0.99\textwidth}
                 \includegraphics[width=6.8in, height=1.32in]{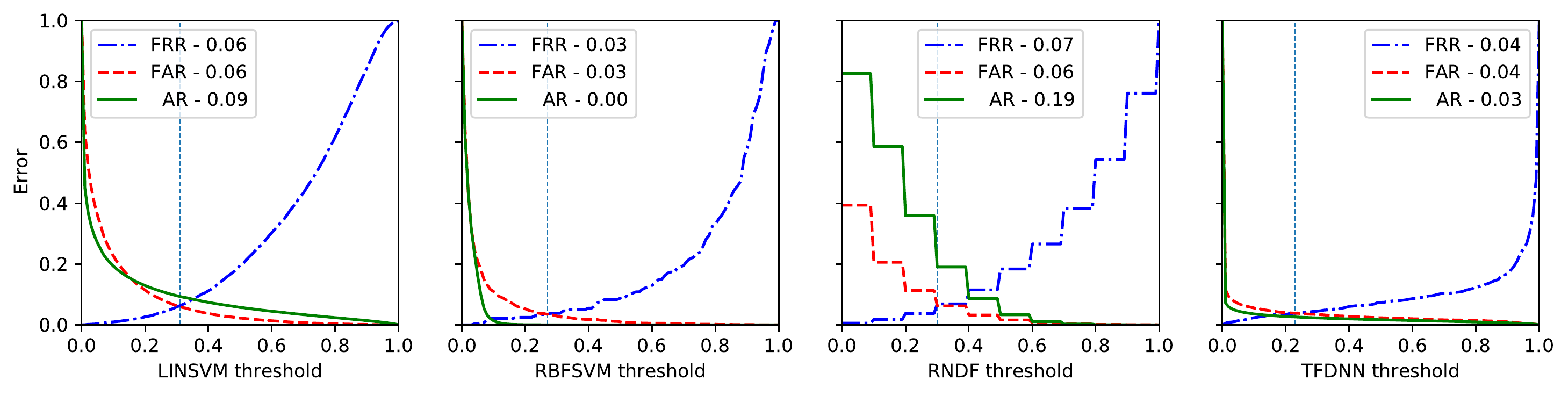}
                 \caption{The performance of \textbf{$v$ABGait} under both Zero-effort (FAR) and Random-vector attack (AR).}
                 \label{v-smartwatch}
         \end{subfigure}
         
         \begin{subfigure}[b]{0.99\textwidth}
                 \includegraphics[width=6.8in, height=1.32in]{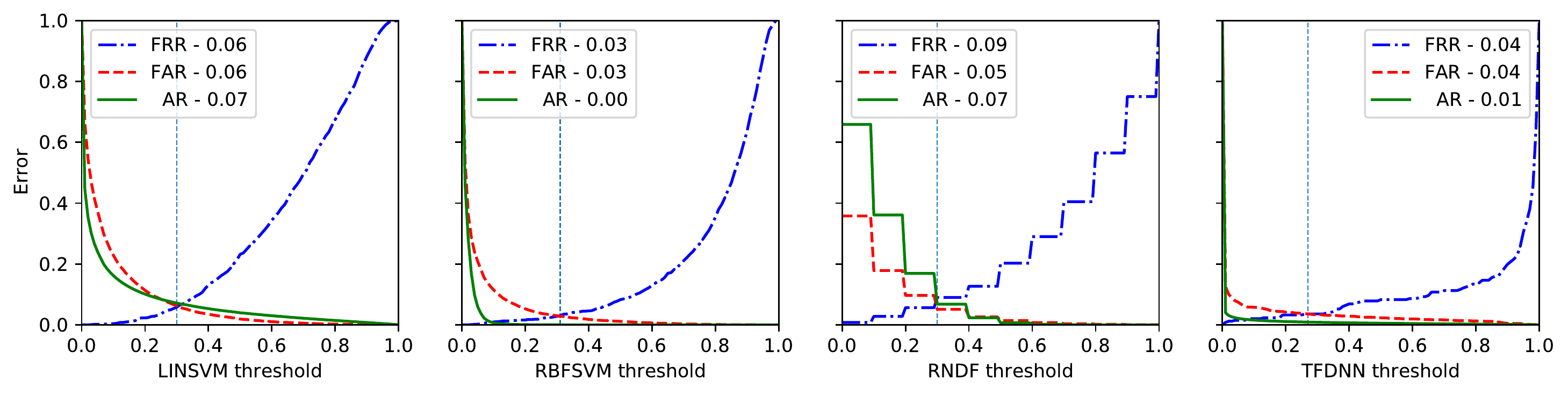}
                 \caption{The performance of \textbf{$\beta$ABGait} under both Zero-effort (FAR) and Random-vector attack (AR).}
                 \label{b-smartwatch}
         \end{subfigure}
         
         \begin{subfigure}[b]{0.99\textwidth}
                 \includegraphics[width=6.8in, height=1.32in]{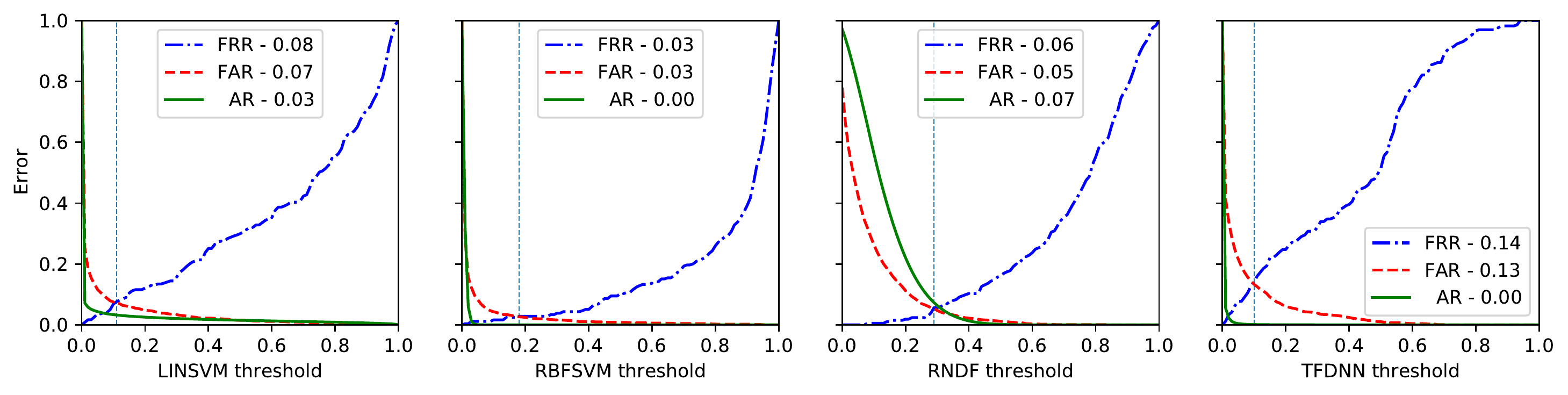}
                 \caption{The performance of \textbf{$i$ABGait} under both Zero-effort (FAR) and Random-vector attack (AR).}
                 \label{i-smartwatch}
         \end{subfigure}
    \caption{The performance of different \textit{ABGait} under different attack scenarios on the \textbf{Watch dataset} \cite{KumarArm}.}
    \label{WatchResults}
\end{figure*} 

The underlying structure of CTGAN that uses fully connected networks in generator $\mathbb{G}$ and critic $\mathbb{C}$ to capture all possible correlations between columns. The network structure of $\mathbb{G}(z, cond)$ adapted from \cite{CTGANOriginal} and reproduced below:  

$
\begin{cases} 
 h_0 = z \oplus cond \\
 h_1 = h_0 \oplus \text{ReLU (BN(FC}_{|cond|+|z| \to 256} (h_0))) \\
 h_2 = h_1 \oplus \text{ReLU (BN(FC}_{|cond|+|z|+256 \to 256} (h_1))) \\
 \hat{\alpha_i} = \text{tanh(FC}_{|cond|+z|+512 \to 1} (h_2)) \text{    }  1 \leq i \leq N_{c} \\
 \hat{\beta_i} = \text{gumbel}_{0.2}\text{(FC}_{|cond|+|z|+512 \to m_i} (h_2)) \text{    } 1 \leq i \leq N_{c} \\
 \hat{d} = \text{gumbel}_{0.2}\text{(FC}_{|cond|+|z|+512 \to D_i} (h_2))  \text{    }  1 \leq i \leq N_{d} \\
\end{cases}
$ 
 
The network architecture for critic $\mathbb{C}$ is below \cite{CTGANOriginal}: 

$
\begin{cases} 
h_0 = r_1 \oplus ... \oplus r_{k} \oplus cond_1,...,cond_k \\
h_1 = \text{drop(leaky}_{0.2}\text{(FC}_{k|r|+10|cond|\to 256}(h_0))\\
h_2 = \text{drop(leaky}_{0.2}\text{(FC}_{256 \to 256}(h_1)\\
\mathbb{C(.)} = \text{FC}_{256 \ to 1} (h_2)
\end{cases}
$

The networks were trained using Wasserstein GAN loss \cite{WGANGP} with Adam optimizer and learning rate $2 \times 10^{-4}$ at $300$ epochs as described in \cite{CTGANOriginal}. 

\subsection{Authentication Frameworks} The training framework is demonstrated in Figure \ref{AdversarialISGait}. As illustrated, the Train and Validate process has three incoming arrows, one for each ABGait implementation strategy, i.e., Vanilla, $\beta$NGen, and $i$CTGAN, resulting in $v$ABGait, $\beta$ABGait, and $i$ABGait, respectively. In short, these frameworks differ in how the impostor samples were obtained for training the authentication models. 

\subsection{Testing Environments} Once trained, each variant of ABGait was tested for the genuine pass and impostor fail tests. The genuine pass test was conducted on the feature vectors extracted from genuine user data one by one with the expectation that the feature vectors would be classified as genuine. In contrast, the impostor fail test included two different experimental setups, viz., Zero-effort attacks and random-vector attacks are described below:

\subsubsection{Zero-effort attack}
The zero-effort attack is a commonly used testing environment. In this environment, feature vectors of users other than the genuine user are considered impostor samples. In other words, for testing the authentication model of user $U_i$, we use the data collected from all the users but $U_i$. The impostors make no active effort or receive training to copy or imitate $U_i$, therefore the name zero-effort attack environment. 

\subsubsection{Random-vector attack}
Before ABGait gets deployed in a critical environment, it must be tested rigorously, especially for active adversarial attempts, such as the random-vector attack \cite{Zhao2020}. To launch the random-vector attack, the attacker needs to know the length of the feature vector and the range of values it contains for each feature. The proposers \cite{Zhao2020} of the random-vector attack believe that it is possible to acquire this information by probing the ABGait API, which would be rather easily accessible. The attacker will try a million randomly generated feature vectors to find the kind of feature vectors accepted by the ABGait API. It is similar to trying randomly generated combinations of passwords. 

\subsection{Performance Evaluation}
To evaluate the performance of each of the ABGait frameworks, we use the False Accept Rate (FAR), False Reject Rate (FRR), Half Total Error Rates (HTER), Receiver Operating Characteristic (ROC) curves, and Acceptance Region (AR) \cite{Zhao2020}. These metrics are defined as follows, where FA means False Accepts, TR means True Rejects, FR means False Rejects, and TA means True Accepts: \\

$
\begin{cases} 
FAR = FA/(FA+TR), \\ 
FRR = FR/(FR+TA),  \\
HTER = (FAR+FRR)/2 \\
A_{R} := \{{x \in  \mathbb{I}^{n} : R(x) = genuine}\} \\
\end{cases}
$
\\

$\mathbb{I}^{n}$ is the $n$-dimensional unit cube that represents a min-maxed normalized feature space, $x$ is a uniformly random input, and $R$ is the model that takes a feature vector $x$ and outputs a predicted label of $genuine$ or $impostor$. Zhao \etal \cite{Zhao2020} argues that AR is directly correlated with the success of random vector attacks. The greater the AR is, the higher the probability of a random vector attack's success. In other words, the lower the AR is, the better the ABGait. 

\begin{figure*}[htp]
        \begin{subfigure}[b]{0.3\textwidth}
                 \centering
                 \includegraphics[width=2in, height=1.43in]{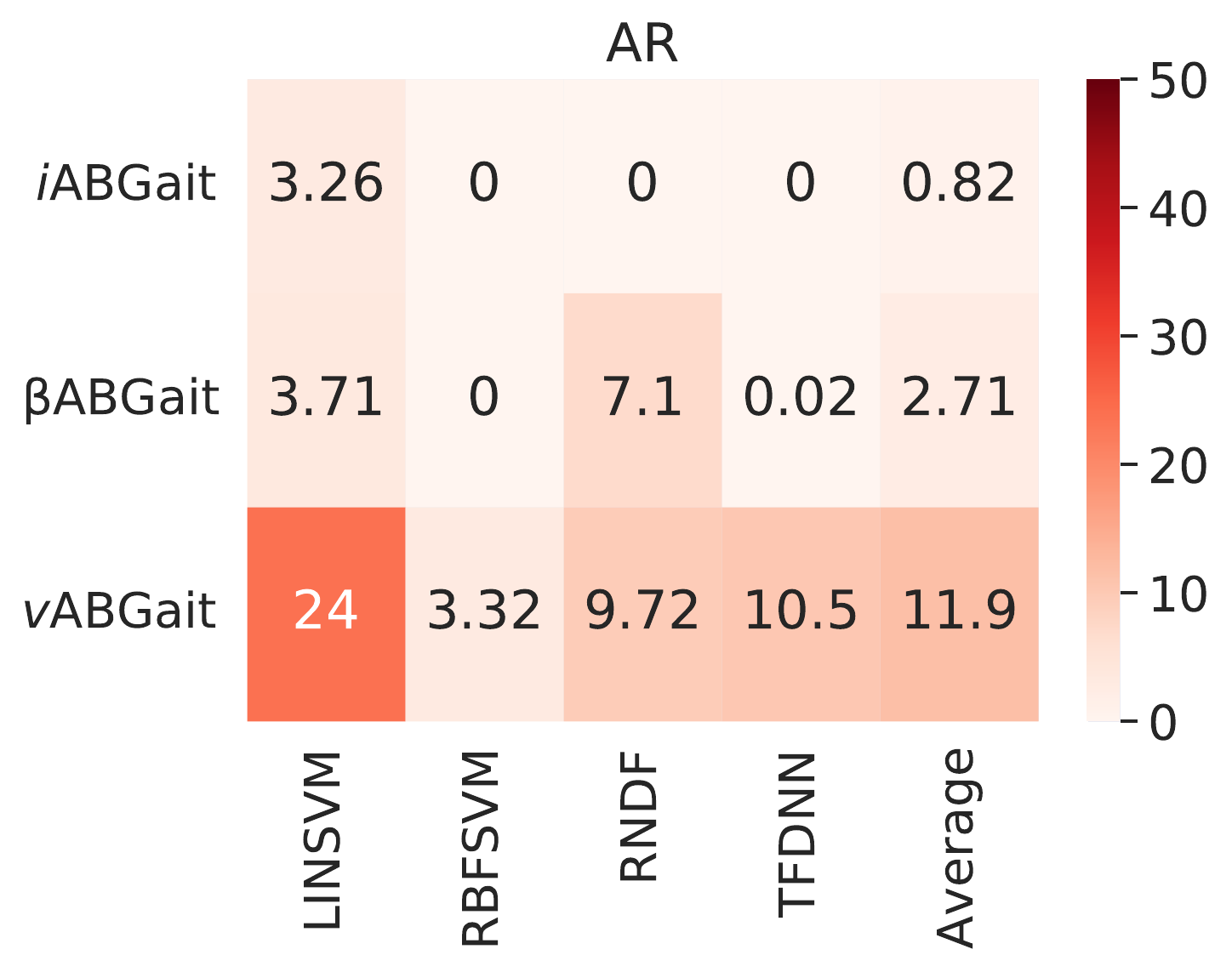}
                 \caption{HAR dataset}
                 \label{heatarhar}
         \end{subfigure} \hfill
         \begin{subfigure}[b]{0.3\textwidth}
                 \centering
                 \includegraphics[width=2in, height=1.43in]{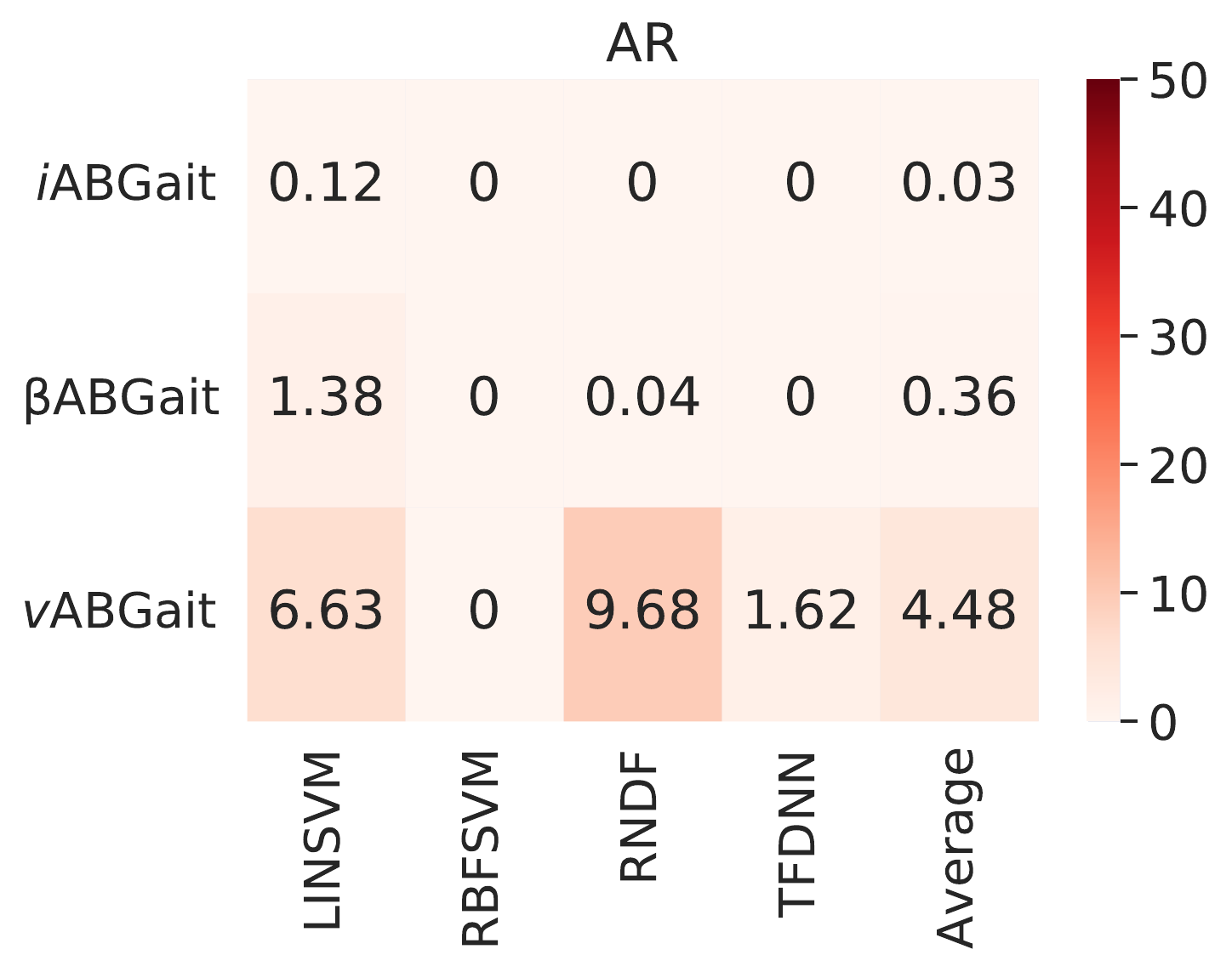}
                 \caption{Phone dataset}
                 \label{heatarphone}
         \end{subfigure}\hfill
         \begin{subfigure}[b]{0.3\textwidth}
                 \centering
                 \includegraphics[width=2in, height=1.43in]{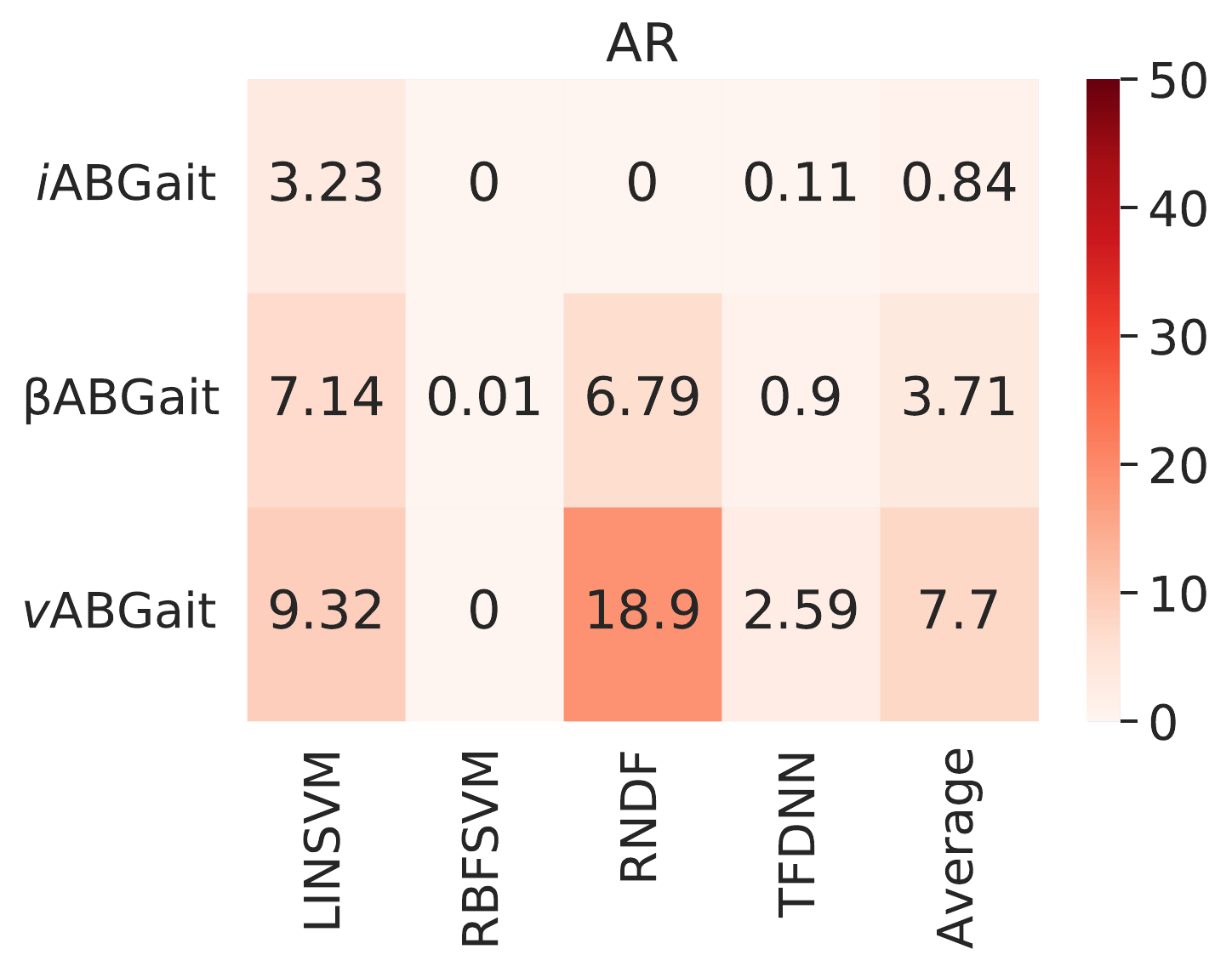}
                 \caption{Watch dataset}
                 \label{heatarwatch}
         \end{subfigure}
         \begin{subfigure}[b]{0.3\textwidth}
                 \centering
                 \includegraphics[width=2in, height=1.43in]{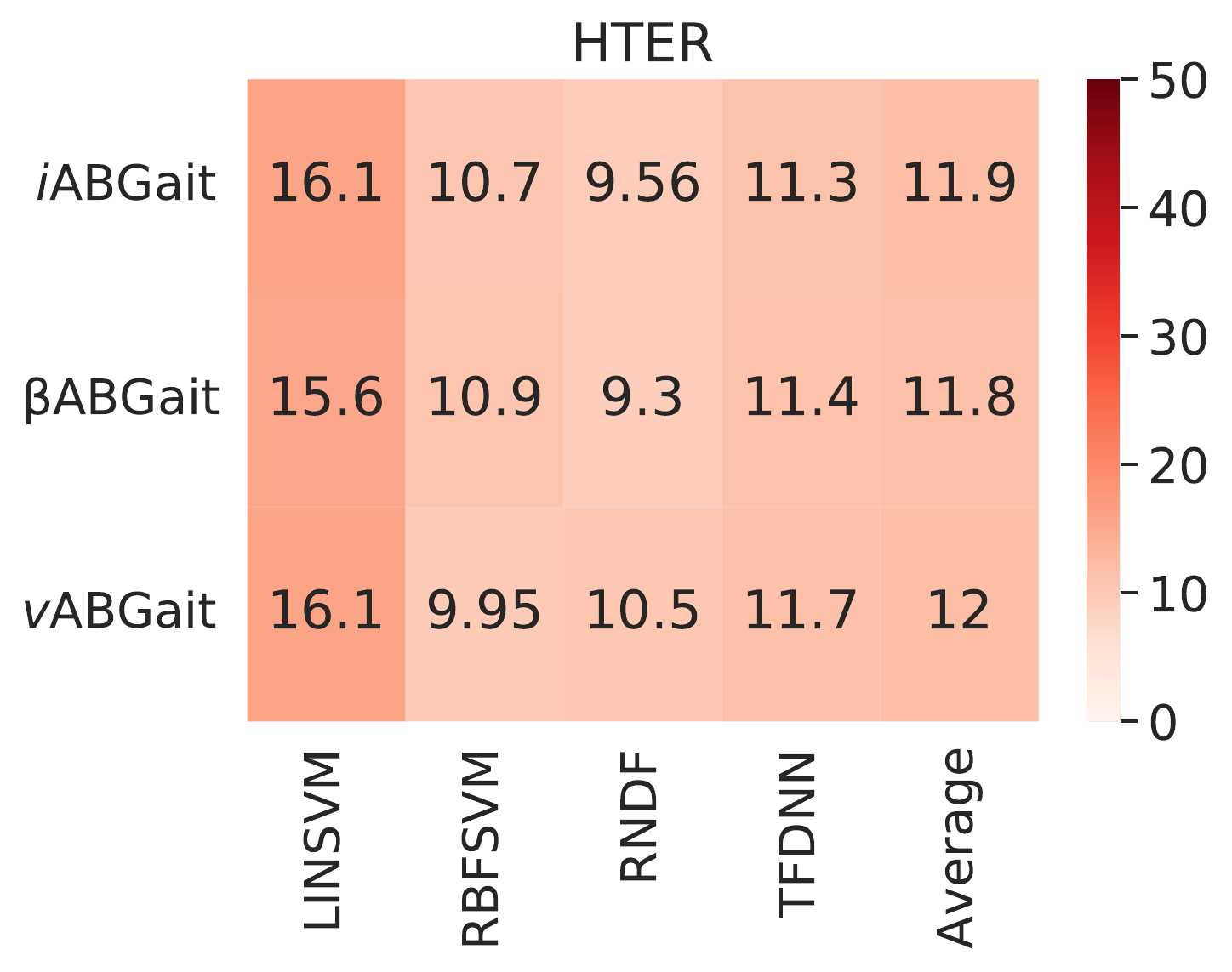}
                 \caption{HAR dataset}
                 \label{heathterhar}
         \end{subfigure} \hfill
         \begin{subfigure}[b]{0.3\textwidth}
                 \centering
                 \includegraphics[width=2in, height=1.43in]{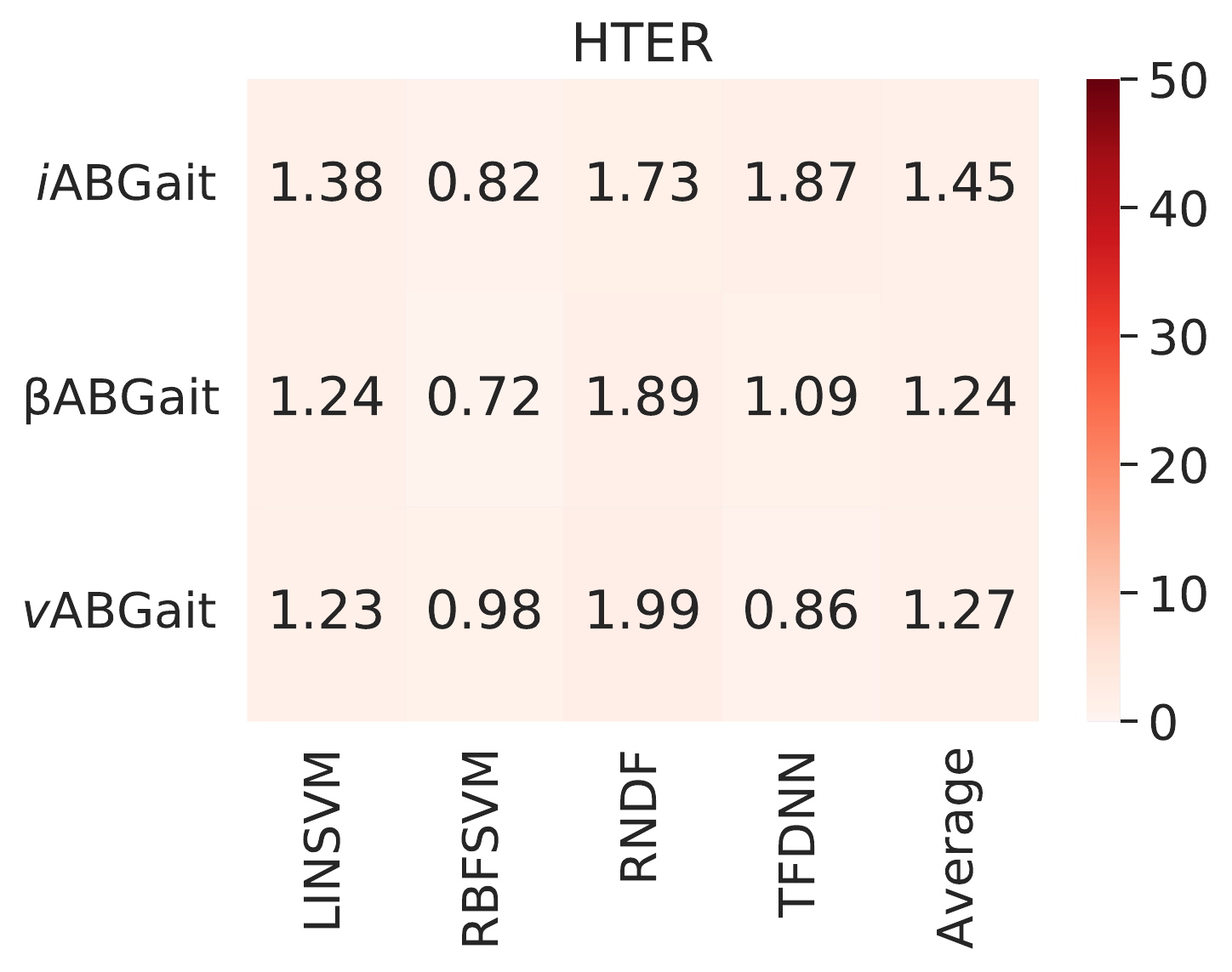}
                 \caption{Phone dataset}
                 \label{heathterphone}
         \end{subfigure}\hfill
         \begin{subfigure}[b]{0.3\textwidth}
                 \centering
                 \includegraphics[width=2in, height=1.43in]{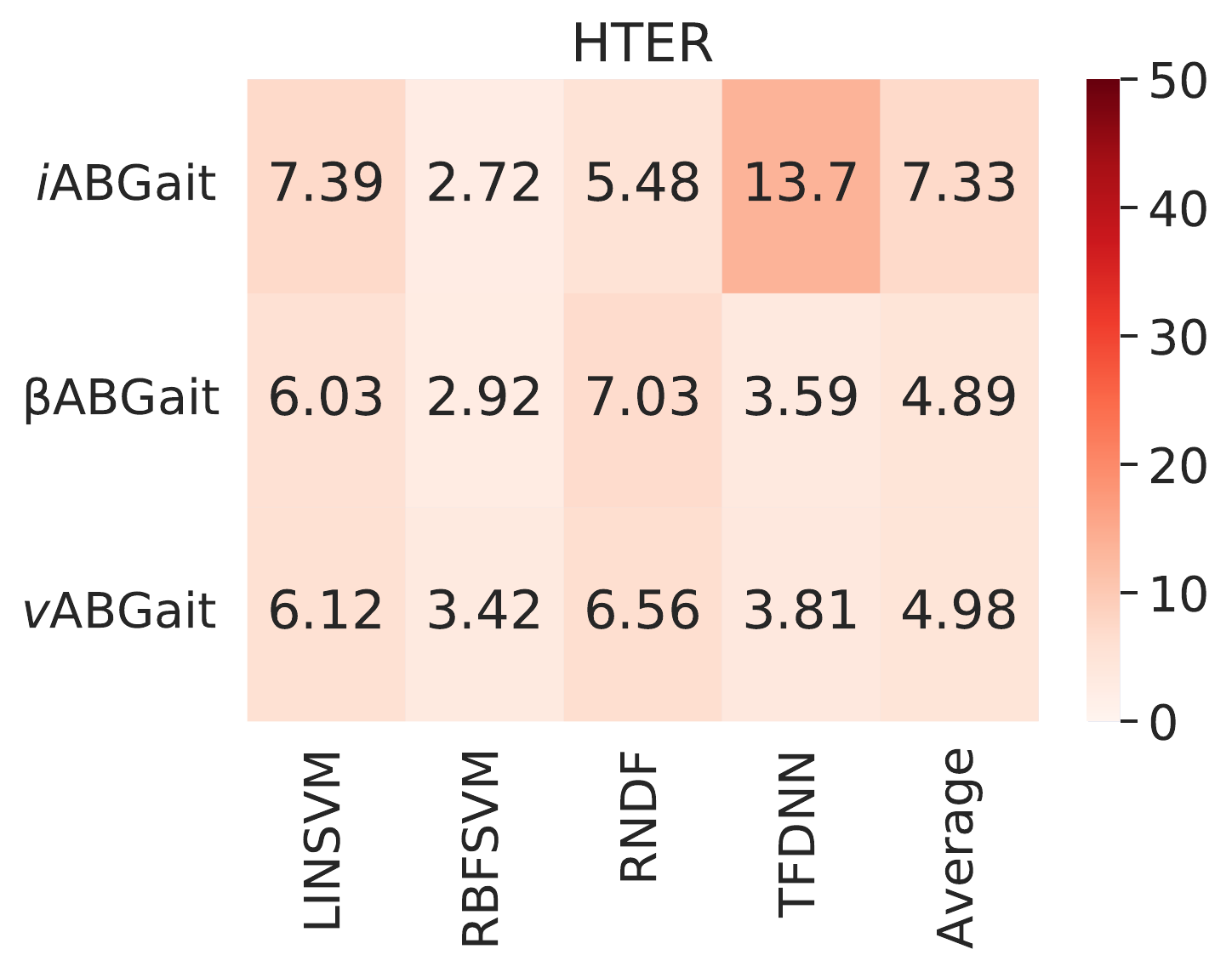}
                 \caption{Watch dataset}
                 \label{heathterwatch}
         \end{subfigure}
\caption{Summary of the HTER and AR obtained by the proposed mitigation technique compared to that of the existing one \cite{Zhao2020} and without them. It is evident from these heatmaps that mitigation based on $i$CTGAN is more effective than the method proposed in \cite{Zhao2020}.}
\label{EntireResults}
\end{figure*}

\section{Results and Discussion}
\label{ResultsDiscussion}
Figures \ref{HARResults}, \ref{PhoneResults}, and \ref{WatchResults} present the ROCs indicating FAR, FRR, and AR for $v$ABGait, $\beta$ABGait, and $i$ABGait. The average ARs for $v$ABGait, $\beta$ABGait, and $i$ABGait are $12\%$, $1\%$, and $.75\%$, respectively on the HAR dataset. Similarly, the average ARs for $v$ABGait, $\beta$ABGait, and $i$ABGait are $4.75\%$, $0.25\%$, and $0\%$, respectively, on the Phone dataset. Finally, the average ARs for $v$ABGait, $\beta$ABGait, and $i$ABGait are $7.75\%$, $3.75\%$, and $2.5\%$, respectively, on the Watch dataset. The high ARs across the dataset for $v$ABGait indicate that the common implementation of ABGait is vulnerable to random-vector attacks. This observation aligns with the conclusion of \cite{Zhao2020}, but on multiple datasets. Moreover, the superiority of $i$ABGait over $\beta$ABGait is evident across the three datasets as it offers more resilience against the random-vector attack.


Although $\beta$ABGait and $i$ABGait implementations showed robustness against the random-vector attack, it is essential that both achieved comparable HTER to $v$ABGait while achieving lower ARs. Thus, we summarize the AR and HTER using heatmaps in Figure \ref{EntireResults}. 

The AR for $v$ABGait ranges between $3.32\%-24\%$, $0-9.68\%$, and $0-18.9\%$ for HAR, Phone, and Watch datasets, respectively. On the other hand, $\beta$ABGait AR ranges between $0-7.1\%$, $0-1.38\%$, and $0.01-7.14\%$ for HAR, Phone, and Watch datasets, respectively. Likewise, the AR obtained by $i$ABGait ranges between $0-3.26\%$, $0-0.12\%$, and $0.0-3.23\%$, respectively, for HAR, Phone, and Watch datasets. The superiority of $i$ABGait and $\beta$ABGait in terms of resilience while maintaining comparable HTERs obtained by $v$ABGait and $\beta$ABGait, is evident from these numbers. 

Figures \ref{heathterhar}, \ref{heathterphone}, and \ref{heathterwatch} show that all three implementations of ABGait i.e. $v$ABGait, $\beta$ABGait, and $i$ABGait achieve comparable performance without random-vector attack taken into consideration. An outlying observation is TFDNN which achieved significantly high $(13.7\%)$ HTER for $i$ABGait compared to the rest of the classifiers on the Watch dataset. We plan to investigate the reason behind the same in the future. In addition, a previous study \cite{RandomAttackNoEffectOnDistanceBased} has reported that the random-vector attacks are ineffective on distance-based classifiers. We will investigate the same in the future on multiple datasets with multiple distance-based classification algorithms. Even though the distance-based classifiers achieve inferior performance than the machine learning-based classifiers \cite{DistanceBasedVsMachineLearning}, they should be developed further if they are immune to random-vector attacks. 


This work reaffirms that increasing the variance of the impostor samples either via $\beta$ noise injection or with the proposed $i$CTGAN helps mitigate random vector attacks to a good extent. The conclusion holds for multiple datasets (watch and phone) and classification algorithms. The proposed technique $i$CTGAN achieved superior results than $\beta$ noise injection. We aim to investigate whether the $i$CTGAN-based mitigation process applies to authentication systems that utilize behavioral patterns other than accelerometer-based gait. In addition, we would investigate how effective the Random-vector attack and the proposed countermeasures are on multi-modal gait biometrics that uses signals from multiple (phone and watch) devices simultaneously \cite{Shrestha2016}. 

\section{Conclusion and Future Work}
\label{Conclusion} 
We studied the impact of random-vector attacks on ABGait using three different datasets. The results reaffirmed that random-vector attacks significantly degrade the performance of ABGait. Furthermore, we tested the effectiveness of $\beta$ noise injection-based mitigation techniques on the three datasets. The technique reduced the AR but left some possibility for improvement. We proposed a novel CTGAN-based mitigation technique and evaluated its effectiveness. The proposed mitigation technique achieved superior results than the $\beta$ noise injection-based technique. We conclude that both $\beta$ noise injection and CTGAN-based impostor oversampling mitigate the impact of the random-vector attack to a good extent. The effectiveness of the fusion of both techniques would be a good topic for future research. In addition, it remains to be investigated whether these techniques will be effective against more sophisticated attacks such as \textit{imitation-based} attacks on ABGait. Moreover, it would be interesting to investigate whether the proposed technique is scalable to authentication systems other than ABGait.   




\section{Acknowledgment}
We are grateful to the anonymous reviewers for their insightful feedback and comments on the paper. We are also thankful to Prof. David Wonnacott, who reviewed the early form of this work at Haverford College. 

{\small
\bibliographystyle{unsrt}
\bibliography{references}
}

\end{document}